\documentclass{article}
\usepackage{authblk}
\usepackage{geometry}
\usepackage{graphicx}
\usepackage{caption}
\usepackage{xcolor}
\usepackage{float}
\usepackage{soul}
\graphicspath{{./}}
\usepackage{hyperref}
\usepackage{cancel}

\usepackage{cancel}
\usepackage{chngcntr}
\counterwithin*{equation}{section}
\counterwithin*{equation}{subsection}

\usepackage [english]{babel}
\usepackage [autostyle, english = american]{csquotes}
\MakeOuterQuote{"}
\usepackage{relsize}
\usepackage{multirow}
\usepackage{subcaption}
\usepackage{bigints}
\usepackage{amsmath, amssymb, physics, bm}
\usepackage{tabularx}
\usepackage{booktabs}
\usepackage{orcidlink}
\usepackage{placeins}
\usepackage{cleveref}
\title{Canonical regularization of the stationary Coulomb problem and an Aufbau-like spectral ordering}

\author[1]{Anand Aruna Kumar\,\orcidlink{0000-0001-6148-2777}}

\affil[1]{Research Engineer, IBM Research, Albany, NY, USA\par
	\href{mailto:anand.aruna.kumar@ibm.com}{anand.aruna.kumar@ibm.com}}

\begin{document}
	\maketitle

\begin{abstract}
 \noindent	The stationary hydrogen atom has Coulomb degeneracy across orbital levels, whereas the Aufbau/Madelung ordering is an empirical, many--electron rule established in atomic physics. We examine the hydrogen atom through a regularized de Broglie--Bohm representation, in which stationary amplitude--current constraints generate separable Sturm--Liouville branches. In this formulation, the radial, orbital, and magnetic sectors acquire canonical Langer--like inverse--square corrections. The modified boundary--value problems allow analytical solutions and produce a hydrogen--like spectrum with regularized radial and angular indices. Consequently, radial Coulomb quantization acquires an orbital--dependent shift, lifting the Coulomb degeneracy and producing a spectral ordering that follows the Aufbau/Madelung sequence.
 \vspace{0.5pc}
 
 \noindent	On this basis, we construct the ordering of the regularized de Broglie--Bohm states and show that the spectral structure retains the standard degenerate Rydberg sequence in the \(l=0\) sector. The separated amplitudes are represented by generalized special function branches, including the associated Laguerre,  Legendre, and Bessel functions with non--integral parameters arising from regularized separation. Therefore, the treatment is intended as an analytical examination of spectral ordering in a regularized one-center Coulomb problem rather than as a replacement for the many--electron atomic structure theory.
 \vspace{0.5pc}
 
	 
\end{abstract}

\section{Introduction}
\vspace{0.5pc}

The Schr\"{o}dinger wavefunction written in the amplitude--phase form,
$\Psi=R\exp(iS/\hbar)$, separates the stationary dynamics into coupled amplitude and phase equations in the de Broglie--Bohm representation \cite{Holland1993}. Under stationary conditions, the zero--current branch of the amplitude equation can be reduced to the Ermakov--Pinney form \cite{Ermakov1880}, and closed--form solutions of the associated coupled equations are available for elementary systems \cite{eins, Schuch2}. However, for finite--current branches, the amplitude and momentum fields remain coupled under the stationary current condition. This coupling can generate singular points in a separated coordinate representation, which requires an additional regularization condition.
\vspace{0.5pc}

\noindent A recent study by the author and collaborators \cite{AS} introduced a constrained Fisher--variational method to regularize amplitude--phase singularities. In the separated representation, regularization is expressed through the dynamical variables ($p_i,q_i$), which satisfy the canonical relation
\begin{equation}
	p_i q_i = \frac{\hbar}{2},
\end{equation}
for each separable coordinate sector. A comprehensive list of the corresponding amplitude transformation functions, obtained through Liouville scaling of the Sturm--Liouville equations, was provided for stationary systems in Ref.~\cite{AAKErmakov}. The resulting component equations acquire Langer--like inverse--square corrections \cite{Langer}, while retaining a separable Sturm--Liouville structure. Therefore the amplitude functions remain within the special function families associated with the corresponding differential equations. However, imposing terminating conditions on the associated hypergeometric branches can shift the resulting energy quantization.
\vspace{0.5pc}

\noindent The shifted inverse--square terms also connect the present construction to the Langer--like corrections in radial quantum problems. In the WKB and uniform--asymptotic treatments, the Langer modification refines the radial Coulomb quantization by correcting the centrifugal behavior near the second--order pole at the origin. For example, Li, Zhu and Wang~\cite{lizhuwang} related this correction to the finiteness of the corresponding error--control function and applied it explicitly to a hydrogen atom. In the present work, an analogous inverse--square shift arises instead from canonical regularization of the stationary amplitude--phase equations, leading to a Langer--like regularized Coulomb spectrum in a de Broglie--Bohm representation.

\vspace{0.5pc}

\noindent We applied this regularized stationary formulation to the hydrogen atom problem in spherical coordinates. Canonical inverse--square regularization modifies the separated radial, orbital, and axial equations, producing an energy quantization condition that depends on both $n_r$ and $l$. The regularized angular sector also differs from the standard spherical--harmonic construction: the orbital branch is described by associated Legendre functions with non--integral degree and order, whereas the axial branch takes a Bessel function form rather than a purely harmonic factor $e^{i m\phi}$. Thus the separated angular labels are not fixed by the usual spherical--harmonic condition on $Y_{lm}(\theta,\phi)$, allowing the orbital and axial labels to be treated as regularized branches before the minimum hydrogenic labeling scheme is imposed. Thus, separated angular amplitudes are not required to satisfy the usual $SO(3)$ spherical--harmonic closure, and their orbital and axial labels may be treated as regularized branches before the minimum hydrogenic labeling scheme is imposed.
\vspace{0.5pc}

\noindent
The principal observation of this communication is that the resulting regularized hydrogen--like branches can be organized into an Aufbau--like arrangement of states \cite{madelung}. In standard quantum mechanics, the Coulomb energy depends only on the principal quantum number, whereas the ordering relevant to atomic subshell filling is conventionally identified from the empirical spectral structure and many--electron atomic behavior.

\vspace{0.5pc}
\noindent It is important to distinguish the present results from conventional corrections to the hydrogenic spectrum and from many--electron electronic--structure calculations. Relativistic and radiative corrections, including the Lamb shift, lift particular degeneracies and refine spectroscopic level positions. However, they are not ordinarily formulated as a mechanism for generating the Aufbau ordering of orbital sectors. Conversely, many--electron approaches reproduce orbital--energy inversions and configuration irregularities through screening, penetration, exchange, correlation, and relativistic effects. Thus, the resulting ordering is a property of the interacting atomic problem, rather than an exact analytical sequence of the one-center Coulomb equation. Several attempts have also been made to rationalize the empirical Madelung or Aufbau ordering beyond its rule--based formulation, including effective screening and penetration arguments, Thomas--Fermi or mean-field estimates~\cite{vmklechkovskii1, vmklechkovskii2}, modified central potentials~\cite{yunostro}, and alternative ordering functions constructed from combinations of principal and orbital quantum numbers. A recent complementary approach by Baez~\cite{Baez} obtains Madelung--type rules from a second--quantized Kepler model based on hidden $SU(2)\times SU(2)$ symmetry and a modified Fock--space Hamiltonian. The present work instead isolates a simpler question: Can a one-center stationary Coulomb problem, written in a canonically regularized de Broglie--Bohm representation, already generate intrinsic splitting among orbital branches?

\vspace{0.5pc}

\noindent
The remainder of this paper is organized as follows. 
Section~\ref{sec:Sec2} formulates the regularized hydrogen problem from amplitude--phase equations in spherical coordinates and obtains the corresponding separated solutions and energy spectrum. Section~\ref{sec:Sec3} develops the mathematical framework for regularized amplitudes and their relationship to the energy quantization that underlies spectral ordering. Section~\ref{sec:Sec4} presents the regularized energy spectrum and compares its ordering with the Aufbau/Madelung sequence. Section~\ref{sec:Sec5} provides the summary and conclusions.

\section{Hydrogen atom in de Broglie--Bohm form}
\label{sec:Sec2}
The stationary wavefunction in the de Broglie--Bohm representation for the amplitude and phase with stationary current conditions was used to obtain component-wise differential equations. The equations are separated into their respective Hamilton--Jacobi energy and current equations and the Sturm--Liouville amplitude regularization for each ($p_i,q_i$) pair is included to obtain modified amplitude solutions for all coordinates. 

\subsection{Stationary amplitude--phase equations in spherical coordinates}
\label{subsec:dbb-spherical}
The wavefunction $\Psi = R\exp(iS/\hbar)$ in spherical form  $H(p_r,p_{\theta},p_{\phi})+\partial S/\partial t =0$ in the de Broglie--Bohm form is given by
\begin{equation}
	\frac{p^2}{2m_p} +V(r) -\frac{\hbar^2}{2m_p}\frac{\nabla^2{R}}{R} = E
\end{equation}
and the stationary phase equation is given by
\begin{equation}
	\nabla\cdot(R^2\nabla{S}) = 0
\end{equation}

\noindent	Here \(p_r,p_\theta,p_\phi\) denote the orthonormal components of
\(\nabla S\) in spherical coordinates,
\[
p_r=\partial_r S,\qquad
p_\theta=\frac{1}{r}\partial_\theta S,\qquad
p_\phi=\frac{1}{r\sin\theta}\partial_\phi S .
\]
\vspace{0.5pc}

\noindent	Explicitly, expressing the componentwise amplitude functions with $R(r,\theta,\phi) = \rho(r)\Theta(\theta)\Phi(\phi)$ and $S(r,\theta,\phi)= S_r(r)+S_{\theta}(\theta) +S_{\phi}(\phi)$ with the respective component momentas as $\vec{p} =\nabla{S}$, we get

\begin{equation}
	\label{eq:coulombhamitoniane}
	\frac{p_r^2}{2m_p}
	+\frac{p_\theta^2}{2m_p}
	+\frac{p_\phi^2}{2m_p}
	-\frac{\hbar^2}{2m_p r^2}
	\left[
	\frac{1}{\rho}\frac{d}{dr}\left(r^2\frac{d\rho}{dr}\right)
	+
	\frac{1}{\Theta\sin\theta}
	\frac{d}{d\theta}
	\left(\sin\theta\frac{d\Theta}{d\theta}\right)
	+
	\frac{1}{\Phi\sin^2\theta}
	\frac{d^2\Phi}{d\phi^2}
	\right]
	+V(r)=E .
\end{equation}

\noindent Thus the metric factors are absorbed into the definition of the
component momenta.
\subsection{Regularized separation of the Coulomb problem}

\vspace{0.5pc}

\noindent	The amplitude and phase are constrained for stationary current, which, upon regularization leads to the following relations between canonical variables and:
\begin{equation}
	\label{eq:canonbohm}
	p_r\cdot r = \frac{\hbar}{2},  \quad p_{\theta}\cdot r sin\theta= \frac{\hbar}{2}, \text  { and } \quad p_{\phi}\cdot r\phi = \frac{\hbar}{2}.
\end{equation}
\noindent Using the above conditions, the regularized Hamilton--Jacobi equation is given by

\begin{equation}
		\label{eq:regcoulomb}
	\frac{\hbar^2}{8m_p r^2}
	+
	\frac{\hbar^2}{8m_p r^2\sin^2\theta}
	+
	\frac{\hbar^2}{8m_p r^2\phi^2}
	-
	\frac{\hbar^2}{2m_p r^2}
	\left[
	\frac{1}{\rho}\frac{d}{dr}\left(r^2\frac{d\rho}{dr}\right)
	+
	\frac{1}{\Theta\sin\theta}\frac{d}{d\theta}
	\left(\sin\theta\frac{d\Theta}{d\theta}\right)
	+
	\frac{1}{\Phi\sin^2\theta}\frac{d^2\Phi}{d\phi^2}
	\right]
	-
	\frac{\alpha}{r}
	=E .
\end{equation}
\vspace{0.5pc}

\noindent We choose a separation scheme analogous to the spherical--harmonic construction, with \((l,m)\equiv(\lambda,\mu)\) at the level of the regularized component equations. It is convenient to write the axial separation constant as \(\mu(\mu+1)\), rather than \(\mu^2\), because it combines with the canonical term to produce the shifted order \(\mu+\tfrac12\) in the angular amplitudes. This choice also simplifies the later comparison between the regularized labels \((\lambda,\mu)\) and the conventional orbital and magnetic notations.
\vspace{0.5pc}

\noindent Here, \(\lambda\) and \(\mu\) are introduced as separation labels for the regularized problem. Their later relation to the usual orbital and magnetic notation is imposed only through a minimum labeling scheme for comparison with hydrogenic states.

\subsubsection{Radial Coulomb branch}

The separated radial differential equation for the Coulomb potential $ V(r)= -\alpha/r$, ($\alpha=Ze^2$) , is given by,
\begin{equation}
	\frac{\hbar^2}{8m_pr^2}-\frac{\hbar^2}{2m_p}\frac{1}{r^2\rho}\frac{d}{dr}\left(r^2\frac{d\rho}{dr}\right)-\frac{\alpha}{r} +\frac{\lambda(\lambda+1)\hbar^2}{2m_pr^2} = E.
\end{equation}
After a bit of rearrangement we get
\begin{equation}
	\label{eq:radialeq}
	\frac{1}{r^2\rho}\frac{d}{dr}\left(r^2\frac{d\rho}{dr}\right)+\frac{2m_p}{\hbar^2}\left(E+\frac{\alpha}{r}\right) -\frac{1}{r^2}\left(\lambda(\lambda+1)+\frac{1}{4}\right)=0,
\end{equation}	
where $\lambda(\lambda+1)$ is the orbital separable constant.
\vspace{0.5pc}

\noindent \noindent  The differential equation for \(\rho(r)\) is a modified Laguerre-type confluent hypergeometric equation, and  ~\eqref{eq:radialeq}'s solution can be written as, \cite{Arfken}

\begin{equation}
	\label{eq:rhorsol}
	\rho_{n_r\lambda}(r)
	=
	\mathcal{N}_{n_r\lambda}\,
	r^{\sqrt{\lambda(\lambda+1)+1/2}-1/2}
	e^{-\kappa r}
	L^{2\sqrt{\lambda(\lambda+1)+1/2}}_{n_r}(2\kappa r).
\end{equation}
\begin{equation}
	\label{eq:kappaE}
	\kappa
	=
	\sqrt{-\frac{2m_pE}{\hbar^2}}
	=
	\frac{m_p\alpha}{
		n_r+\frac12+\sqrt{\lambda(\lambda+1)+\frac12}
	}.
\end{equation}
where $\mathcal{N}_{n_r,\lambda}$ is the normalization constant of the radial function and $n_r$s are non-negative integers. The Langer--like term shifts the associated Laguerre parameter in $L_a^b(2\kappa r)$, where $a=n_r$ is the polynomial degree and $b$ is generally non--integral.

\subsubsection{Orbital angular branch}
The reduction of $\Theta(\theta)$ is explicitly written as
\begin{equation}
	\frac{\hbar^2}{8m_p r^2sin^2\theta}-\frac{\hbar^2}{2m_pr^2}\frac{1}{\Theta\cdot sin\theta} \frac{d}{d \theta} \left(sin{\theta}\frac{d \Theta}{ d \theta}\right) +\frac{\hbar^2\mu(\mu+1)}{2m_pr^2sin^2\theta} =\frac{\hbar^2\lambda(\lambda+1)}{2m_pr^2},
\end{equation}
which, upon an algebraic rearrangement simplifies to a Legendre--type hypergeometric differential equation given by 
\begin{equation}
	\label{eq:theta_reg_simplified}
	\frac{1}{\Theta\cdot sin\theta} \frac{d}{d \theta} \left(sin{\theta}\frac{d \Theta}{ d \theta}\right) +\left(\lambda(\lambda+1) -\frac{\mu(\mu+1)+\tfrac14}{sin^2\theta}\right)=0,
\end{equation}
Note that the choice of $\mu(\mu+1)$ along with the canonical relations, reduces to a perfect square term of $(\mu+\tfrac12)$, which is the degree of the modified Legendre equation under canonical regularization. This is a simplified choice of the constant for the axial separation term.
\vspace{0.5pc}

\noindent Introducing $x=\cos\theta$, Eq.~\eqref{eq:theta_reg_simplified}
reduces to the associated Legendre equation
\begin{equation}
	\label{eq:assoclegdiff}
	(1-x^2)\frac{d^2\Theta}{dx^2}
	-2x\frac{d\Theta}{dx}
	+
	\left[
	\lambda(\lambda+1)-\frac{\tilde{\mu}^2}{1-x^2}
	\right]\Theta(x)=0,
	\qquad
	\tilde{\mu}=\pm\left(\mu+\frac12\right).
\end{equation}
Its general solution may be written as
\begin{equation}
	\Theta(\theta)
	=
	A\, P_\lambda^{\tilde{\mu}}(\cos\theta)
	+
	B\, Q_\lambda^{\tilde{\mu}}(\cos\theta),
\end{equation}
where $P_\lambda^{\tilde{\mu}}$ and $Q_\lambda^{\tilde{\mu}}$ denote the associated Legendre functions, or Ferrers functions on the interval $-1\leq \cos\theta\leq 1$, of non--integral order $\tilde{\mu}=\mu+\tfrac12$. We use the Ferrers function of the first kind, $P_\lambda^{\pm\tilde{\mu}}(\cos\theta)$, as the regular angular branch, while excluding the $Q$-branch from the regular sector. Because the regularized angular equation depends on $\tilde{\mu}^2$, the signs $\pm\tilde{\mu}$ correspond to equivalent local order choices. For the chart representation used here, we consider the following:
\[
P_\lambda^{-\tilde{\mu}}(\cos\theta),
\qquad 0\leq\theta<\frac{\pi}{2},
\]
and use the reflected chart across the equator for
\[
\frac{\pi}{2}\leq\theta\leq\pi .
\]
This chart assignment is a branch convention for representing the angular sector and is consistent with the standard analytic continuation properties of the associated Legendre functions of non--integral degree and order~\cite{Maier}.
\vspace{0.5pc}

\noindent 
On the reference chart, the regular orbital branch is therefore written as
\begin{equation}
	\label{eq:assocleg}
	\boxed{
		\Theta(\theta)\propto P_\lambda^{-{\left(\mu+\tfrac12\right)}}(\cos\theta).
	}
\end{equation}

\subsubsection{Axial angular equation}
The separable differential equation of the $\phi$ coordinate is given by
\begin{equation}
	\frac{\hbar^2}{8m_pr^2\phi^2}-\frac{\hbar^2}{2m_pr^2}\frac{1}{\Phi}\frac{d^2\Phi}{d \phi^2} = \frac{\hbar^2\mu(\mu+1)}{2m_pr^2}   
\end{equation}	
which simplifies to
\begin{equation}
	\frac{1}{\Phi}\frac{d^2\Phi}{d \phi^2}+\left(\mu(\mu+1)-\frac{1}{4\phi^2}\right)=0.
\end{equation}

\noindent In the case of the axial differential equation, we have two solutions based on the parameter $\mu$. For $\mu=0$, the differential equation reduces to a power--law type of dependence, whereas for non--zero $\mu$, the solution for $\Phi(\phi)$ is Bessel type. We take one of the convergent solutions of the Bessel function, $J_\nu(k\phi)$ and thus obtain

\begin{equation}
	\boxed{
		\Phi_\mu^{\rm reg}(\phi)
		\propto
		\begin{cases}
			\phi^{\frac{1+\sqrt2}{2}},
			&
			\mu=0,
			\\[8pt]
			\sqrt{\phi}\,
			J_{1/\sqrt2}\!\left(
			\sqrt{\mu(\mu+1)}\phi
			\right),
			&
			\mu\neq0.
		\end{cases}
	}
	\label{eq:Phi_reg}
\end{equation}
The amplitude wavefunction $\Phi(\phi)$ has a $\sqrt\phi$ pre--factor in its solution owing to regularization. Therefore, appropriate phase sign factors were used in the representation to distinguish the range of $0\leq \phi < 2\pi$.

\vspace{0.5pc}

\noindent 
The functional forms of Eqs.~\eqref{eq:assocleg} and \eqref{eq:Phi_reg} show that the regularized angular sector differs from the standard spherical--harmonic representation of a hydrogen atom. The orbital branch is described by associated Legendre functions with non--integral order, whereas the axial branch is represented by a Bessel type solution with a regularized index. Hence the separated angular labels are not restricted by the usual spherical--harmonic admissibility condition that fixes the allowed $m$ values for a given $l$. In a fully regularized representation, this permits a broader countable family of ($\lambda,\mu$) branches. For comparison with the conventional hydrogenic notation, one may then select a restricted subset of these branches, referred to below as the minimum labeling scheme.
\vspace{0.5pc}

\section{Construction of the regularized energy spectra}
\label{sec:Sec3}	
\noindent In the previous section,  regularized separated equations and their analytical radial, orbital, and axial solutions were derived. Energy quantization follows from the terminating condition imposed on the confluent--hypergeometric branch of the radial Coulomb solution, as in the standard treatment of the hydrogen atom~\cite{Landau3}. In the present case, however, the regularized radial condition yields a spectrum that depends on both the radial quantum number $n_r$ and the orbital separation parameter $\lambda$.

\vspace{0.5pc}

\noindent Section~\ref{sec:sec31} first analyzes the radial spectrum $E_{n_r,\lambda}$. In particular, we examine whether the non--degenerate regularized spectrum contains the usual degenerate Rydberg sequence as a limiting branch. This is obtained by minimizing $E_{n_r,\lambda}$ with respect to $\lambda$, which identifies the reference branch used to introduce the minimum hydrogenic labeling scheme.

\vspace{0.5pc}

\noindent For completeness, radial labeling must also be accompanied by a consistent choice of angular branches entering the total separated wavefunction. Therefore, Section~\ref{sec:sec32}, develops an angular--sector representation. As shown in Section~\ref{sec:Sec2}, the regularized angular functions are no longer ordinary spherical harmonics: the orbital branch is represented by associated Legendre functions with a non--integral degree and order, whereas the axial branch has a Bessel function structure rather than a purely harmonic factor $e^{i\mu\phi}$. The square--root prefactors in the regularized angular amplitudes lead to branch--dependent representations, which are handled through a simple chart construction over the ($\theta-\phi$) domain.~\Cref{sec:sec33,sec:sec34} summarize the admissible mathematical representations of the angular sector and the properties of the regularized amplitudes.

\subsection{The radial minimum energy ordering for the bound states, $E_{n_r,\lambda}$}
\label{sec:sec31} 

\noindent We consider the non--degenerate radial energy $E_{n_r,\lambda}$ obtained from the regularized radial solution in Eq.~\eqref{eq:kappaE}. Solving the definition of $\kappa$ for energy yields
\begin{equation}
	\label{eq:Ereg}
	E_{n_r,\lambda}
	=
	-\frac{m_p\alpha^2}
	{2\hbar^2\left(n_r+\tfrac12+\sqrt{\lambda(\lambda+1)+\tfrac12}\right)^2},
	\qquad n_r=0,1,2,\ldots .
\end{equation}
For comparison, the degenerate hydrogenic Rydberg expression can be written using the same radial label as
\begin{equation}
	\label{eq:EQM}
	E^{\mathrm{QM}}_{n_r}
	=
	-\frac{m_p\alpha^2}{2\hbar^2(n_r+1)^2},
	\qquad n_r=0,1,2,\ldots .
\end{equation}

\vspace{0.5pc}

\noindent Because $m_p$, $\alpha$, and $\hbar$ are constants, the regularized energy depends only on $n_r$ and $\lambda$. At this stage, $\lambda$ remains a separation label of the regularized problem and has not yet been assigned to the conventional orbital label $l$. Therefore, we first identify the minimum branch of $E_{n_r,\lambda}$ for a fixed $n_r$ as follows:
\begin{equation}
	\label{eq:minelambda}
	\frac{\partial E_{n_r,\lambda}}{\partial \lambda}=0,
	\qquad
	2\lambda+1=0,
	\qquad
	\lambda=-\frac12 .
\end{equation}
Substituting this value gives
\begin{equation}
	\label{eq:Emin}
	\boxed{
		E_{n_r,\lambda}^{\min}
		=
		E_{n_r,-1/2}
		=
		-\frac{m_p\alpha^2}{2\hbar^2(n_r+1)^2}.
	}
\end{equation}
This result coincides with the degenerate Rydberg expression (\(R_H\)) in Eq.~\eqref{eq:EQM}. Thus the regularized spectrum contains a standard degenerate sequence as its minimum branch.

\vspace{0.5pc}

\noindent We now assign the minimum value $\lambda=-\tfrac12$ to the $s$-sector ($l=0$) and introduce shifted orbital labeling,
\begin{equation}
	\label{eq:lambdatol}
	\lambda=l-\frac12,
	\qquad
	l=0,1,2,\ldots .
\end{equation}
Together with the angular shift $\tilde{\mu}=\mu+\tfrac12$, this defines the minimum labeling scheme used to compare regularized branches with conventional hydrogenic notation. By rewriting the radial amplitude in terms of $l$, we obtain
\begin{equation}
	\label{eq:regularizednlmu}
	\rho_{n_r,l}(r)
	=
	\mathcal{N}_{n_r,l}\,
	r^{\sqrt{l^2+1/4}-\tfrac12}
	e^{-\kappa r}
	L^{2\sqrt{l^2+1/4}}_{n_r}(2\kappa r).
\end{equation}
The corresponding Liouville--reduced Coulomb spectrum becomes
\begin{equation}
	\label{eq:enrl}
	E_{n_r,l}
	=
	-\frac{R_H}{
		\left[
		n_r+\frac12+\sqrt{l^2+\frac14}
		\right]^2
	},
	\qquad
	n_r=0,1,2,\ldots,\qquad l=0,1,2,\ldots .
\end{equation}

\vspace{0.5pc}

\begin{figure}[!h]
	\centering
	\includegraphics[
	width=0.98\textwidth,
	height=.57\textheight,
	keepaspectratio
	]{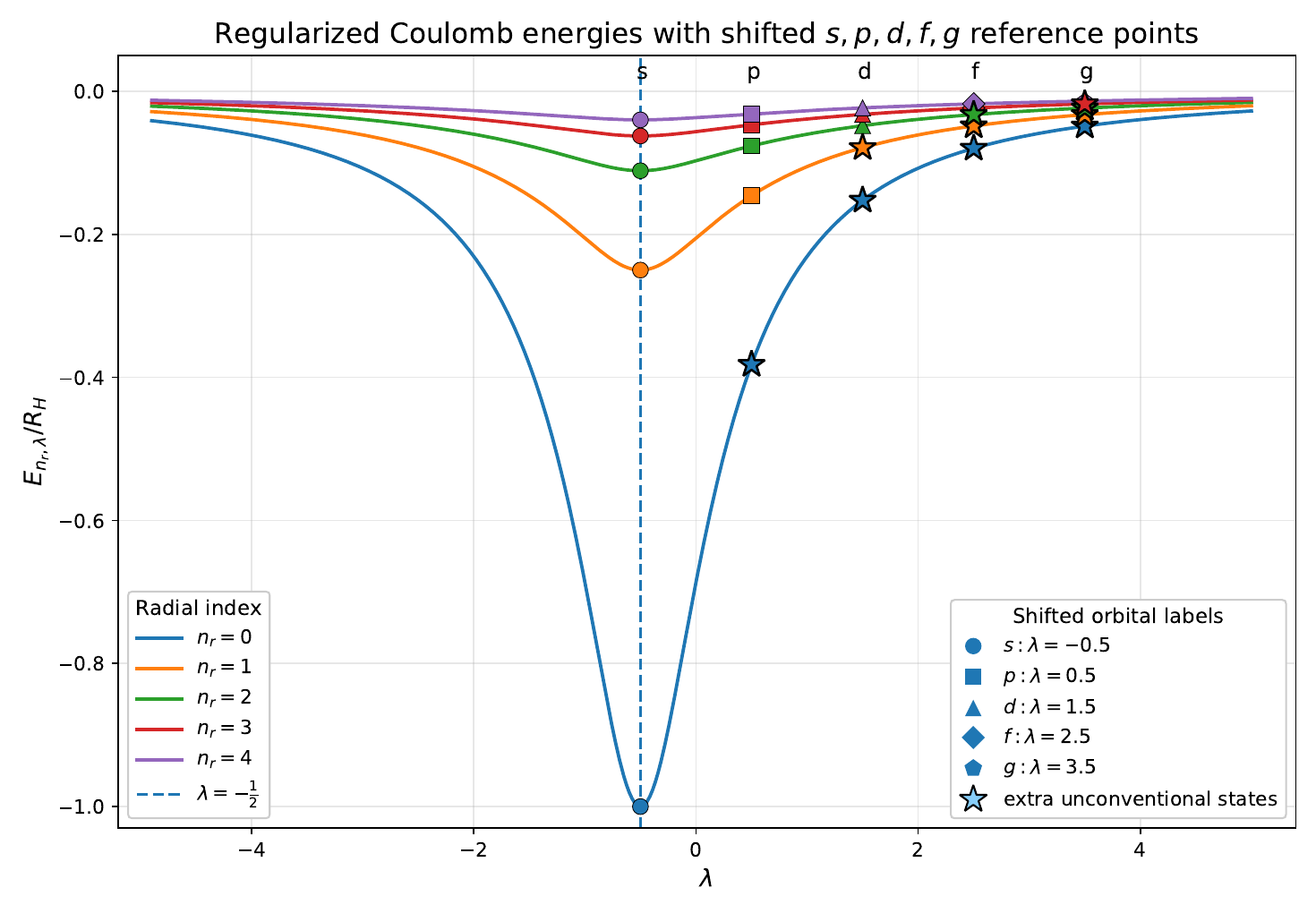}
	\caption{Regularized energy $E_{n_r,\lambda}$ as a function of the separation label $\lambda$.}
	\label{fig:LambdashiftedEnergy}
\end{figure}

\vspace{0.5pc}

\noindent The regularized radial spectrum is shown in Fig.~\ref{fig:LambdashiftedEnergy}. The energy family is symmetric about $\lambda=-1/2$, where the effective radial index reaches its minimum value. Assigning $\lambda=l-\tfrac12$ maps this minimum branch to the $s$-sector and generates the shifted ($s,p,d,f,\ldots$) reference sequence. In the broader regularized representation, the labels ($n_r,l$) need not be restricted by the conventional spherical--harmonic admissibility rules; hence additional branches such as ($1p,1d,2d,\ldots$) can be represented before the minimum comparison scheme is imposed.

\subsection{The $\theta-\phi$ domain angular functions}
\label{sec:sec32}
\noindent It is necessary to examine the solutions of the orbital and axial differential equations to cover the complete angular distribution of regularized amplitudes. 
\vspace{0.5pc}

\noindent
In Section.~\ref{sec:Sec2} we derived the general form of the associated Legendre function  using ~\eqref{eq:assocleg} and ~\eqref{eq:Phi_reg}. The asymptotic form of $P_\lambda^{-\left(\mu+\tfrac12\right)}$ is given by

\begin{equation*}
	{P}_{\lambda}^{-{\left(\mu+\tfrac12\right)}}(cos\theta) \sim \frac{1}{2^{\left(\mu+\tfrac12\right)/2}\Gamma\left(\left(\mu+\tfrac12\right)+1\right)}\left(1-cos\theta\right)^{{\left(\mu+\tfrac12\right)}/{2}}
\end{equation*}
and transformation to $\pi-\theta$ require a different hypergeometric asymptotic form with a non--trivial phase factor. Thus, to cover from $0 \to \pi$, it is convenient to separate into two ranges of $\theta$ at $\theta = \pi/2$, which results in two charts for $\theta$. Furthermore, the Bessel function arguments themselves have a reflection point at $\phi = \pi$, thus to maintain the directionality and single--valued representation of the phase $\phi$ variable, it is necessary to separate them into two charts. 
This is easily achieved through the discrete $\sigma_\theta, \sigma_\phi$ parity operators with values $\pm 1$ which change the sign based on the range of $\theta$ and $\phi$. We compactly represent the angular wavefunction as  
\begin{equation}
	\mathcal{Y}_{\lambda\mu}^{(\sigma_{\theta},\sigma_{\phi})}(\theta,\phi)
	=
	\mathcal{Y}_{\lambda\mu}^0
	\begin{cases}
		{P}_{\lambda}^{-\left(\mu+\tfrac12\right)}(cos\theta)\,{\Phi_{\mu}^{reg}}(\phi),
		
		&

		\substack{
			0\leq\theta<\dfrac{\pi}{2},\\
			\\[8pt]
			0\leq\phi<\pi
		}
		\\[24pt]
		\sigma_{\phi}\,
		{P}_{\lambda}^{-\left(\mu+\tfrac12\right)}(cos\theta)\,{\Phi_{\mu}^{reg}}(2\pi-\phi),
		
		&
		
		\substack{
			0\leq\theta<\dfrac{\pi}{2},\\
			\\[8pt]
			\pi\leq\phi<2\pi
		}
		\\[30pt]
		\sigma_{\theta}\,
		{P}_{\lambda}^{-\left(\mu+\tfrac12\right)}(cos(\pi-\theta))\,{\Phi_{\mu}^{reg}}(\phi),
		
		&
		
		\substack{
			\dfrac{\pi}{2}\leq\theta\leq\pi,\\
			\\[8pt]
			0\leq\phi<\pi
		}
		\\[24pt]
		\sigma_{\theta}\sigma_{\phi}\,
		{P}_{\lambda}^{-\left(\mu+\tfrac12\right)}(cos(\pi-\theta))\,{\Phi_{\mu}^{reg}}(2\pi-\phi),
		
		&
		
		\substack{
			\dfrac{\pi}{2}\leq\theta\leq\pi,\\
			\\[8pt]
			\pi\leq \phi<2\pi.
		}
	\end{cases}
	\label{eq:four_chart_angular_representation}
\end{equation}
The half--open intervals are used only to assign each interface point to a single chart; the interfaces themselves are branch--transfer boundaries and do not affect the radial quantization.
\vspace{0.5pc}

\noindent Thus, it is sufficient to consider the positive branch of the regularized angular representation of the \((\theta,\phi)\) sector in the interval $\left[0,\frac{\pi}{2}\right)\times \left[0,\pi\right)$ and use appropriate parity factors to represent the entire range of angular representation of the system. The positive regular branch function is given by
\begin{equation}
	\label{eq:regylm}
	\mathcal{Y}_{\lambda,{\left(\mu+\tfrac12\right)}}^{(\mathrm{reg+})}(\theta,\phi)
	=
	\mathcal{Y}^0_{\lambda,\left|\left(\mu+\tfrac12\right)\right|}\,
	P_\lambda^{-\left(\mu+\tfrac12\right)}(\cos\theta)\,
	\Phi^{reg}_{\mu}(\phi).
\end{equation}
where $\mathcal{Y}^0_{\lambda,\left|\left(\mu+\tfrac12\right)\right|}$ is the normalization constant.
\vspace{0.5pc}

\noindent Further, from the non--degenerate energy labeling convention, we rewrite the angular wavefunction as
\begin{equation}
	\mathcal{Y}_{l-1/2,{\left(\mu+\tfrac12\right)}}^{(\mathrm{reg+})}(\theta,\phi)
	=
	\mathcal{Y}^0_{l-1/2,\left|\left(\mu+\tfrac12\right)\right|}\,
	P_{l-1/2}^{-\left(\mu+\tfrac12\right)}(\cos\theta)\,
	\Phi^{reg}_{\mu}(\phi).
\end{equation}
\noindent Although the chart--transfer labels $\sigma_{\theta}$ and $\sigma_{\phi}$  specify the polar and azimuthal branch continuations respectively, the signed angular amplitude depends only on the local product on each chart. Thus, the four--chart atlas reduces, at the wavefunction sign level, to two net parity classes while retaining distinct polar and azimuthal chart origins.

\vspace{0.5pc}

\noindent The angular labels were selected only at the level of the comparison scheme. In particular, $\mu$ is assigned integer values such that the shifted order $\mu+\tfrac12$ can be compared with conventional magnetic labeling. This assignment is not imposed by spherical--harmonic $SO(3)$ closure, but is introduced as part of the minimum labeling scheme used to compare the regularized angular branches with hydrogenic notation.

\subsection{Restricted angular branch and persistence of the Aufbau-like ordering}
\label{sec:sec33}
\noindent 
It is useful to distinguish the full canonical regularization of the separated Coulomb problem from a more restrictive radial regularization. In the construction above, the canonical condition is imposed on the radial, orbital, and axial separated sectors. This leads to a generalized angular representation involving associated Legendre functions of non-integral order and an axial Bessel-type branch. However, the spectral ordering mechanism can already be identified before applying canonical regularization to the angular variables.
\vspace{0.5pc}

\noindent
The angular sector may then be interpreted as part of the same shifted separation structure, while not being responsible for the energy ordering itself. Unlike the radial Coulomb equation, where the second-order pole at the origin fixes the role of the Langer-like correction in the radial quantization, the angular domains are not tied to the same singular quantization mechanism. Once the shifted orbital identification
\[
\lambda=l-\frac12
\]
is used, the polar angular equation naturally admits associated Legendre branches of half-integral degree,
\[
\Theta(\theta)\sim P^m_{l-1/2}(\cos\theta),
\]
rather than the ordinary spherical harmonics \(Y_{lm}\). Thus the shifted radial labeling that produces the Aufbau-like ordering also points to a corresponding half-integral angular representation.
\vspace{0.5pc}

\noindent
The axial sector may be treated at different levels depending on the desired functional realization. One may retain a harmonic axial factor such as \(e^{im\phi}\), or allow shifted phase branches, while the fully canonical angular regularization produces further non-integral-order Legendre functions together with regularized axial amplitudes. These alternatives change the detailed angular geometry of the separated amplitudes, but they do not alter the radial energy-ordering mechanism emphasized here.
\vspace{0.5pc}

\noindent
Accordingly, the present paper focuses on the radial Sturm--Liouville/Laguerre origin of the shifted Coulomb spectrum and its Aufbau-like ordering. A fuller synthesis of the associated angular functions, including half-integral-degree branches and possible connections with toroidal or ellipsoidal harmonic structures, will be treated separately.

\subsection{Mathematical properties of regularized amplitudes}
\label{sec:sec34}
\noindent 
From the regularized amplitudes in Eq.~\eqref{eq:regularizednlmu}, the three mathematical features are as follows:
\begin{enumerate}
	\item The radial sector contains an associated Laguerre factor with a generally non--integral associated parameter, except in limiting cases where the regularized index reduces to its integral value. The function remains well--defined through the confluent--hypergeometric or Kummer representation \cite{handbook}.
	
	\item The orbital sector involves associated Legendre functions, the degrees and orders of which are not restricted to integral spherical--harmonic values. The non--integral extension is standard, and the present branch assignment is used only to retain a comparison with conventional orbital labels.
	
	\item The axial sector is represented by Bessel functions of fixed non--integral order \(1/\sqrt{2}\), whereas the separation parameter enters through the effective axial wavenumber \(k_\mu=\sqrt{\mu(\mu+1)}\) for \(\mu\neq 0\).
	
\end{enumerate}
Thus, regularized amplitudes are mathematically admissible as generalized special function branches, and their labels are selected to preserve the comparison with the usual hydrogenic notation.
\vspace{0.5pc}

\section{Arrangement of the regularized energy spectrum, $E_{n_r,l}$, and the Aufbau principle}
\label{sec:Sec4}

\begin{figure}[h!]
	\centering
	\begin{subfigure}[t]{0.7\textwidth}
		\centering
		\includegraphics[
		width=\textwidth,
		keepaspectratio
		]{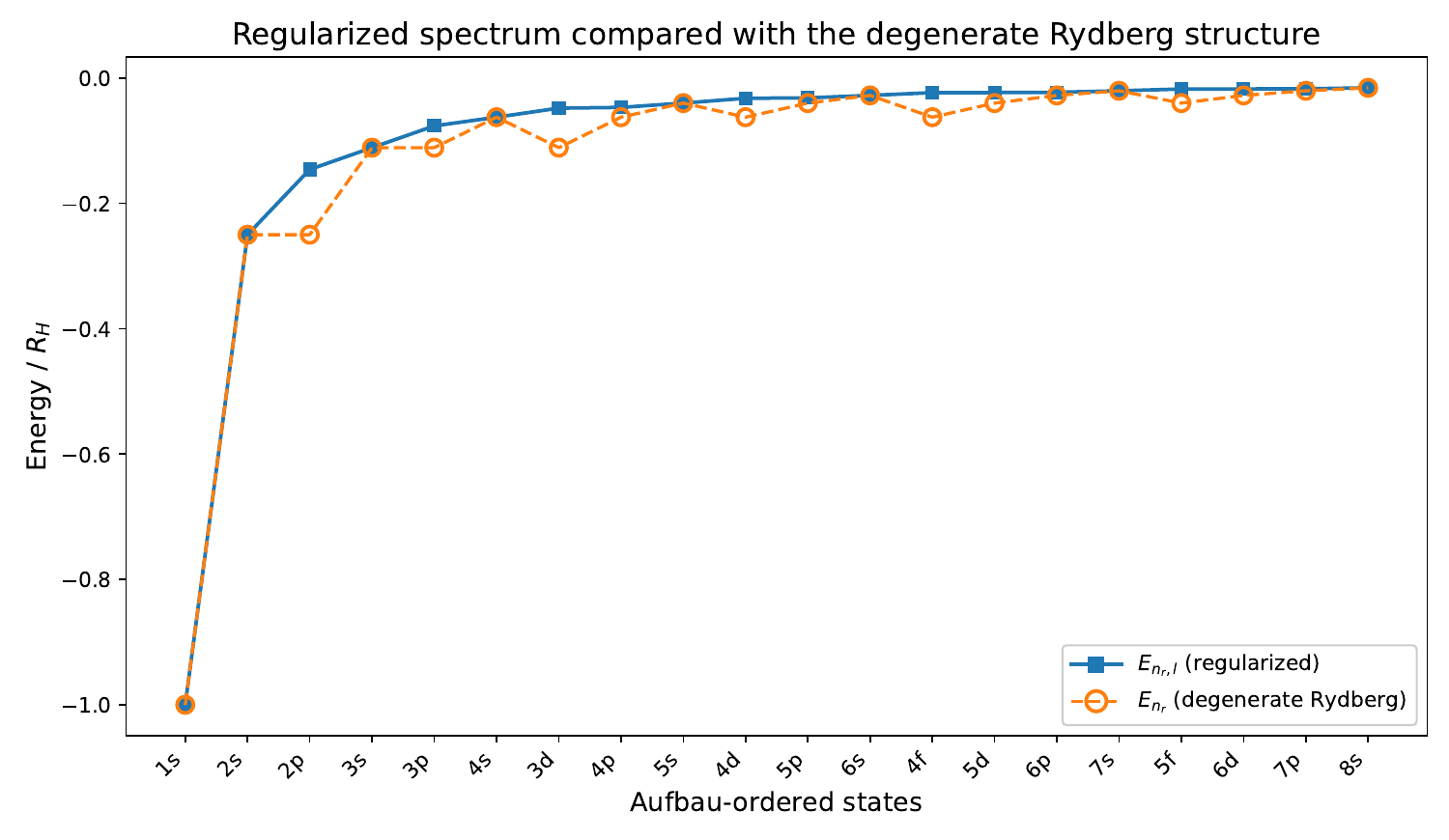}
		\caption{Full Aufbau-ordered sequence from $1s$ -- $8s$.}
		\label{fig:Aufbau_full_overlap}
	\end{subfigure}
	\vspace{0.5pc}
	\begin{subfigure}[t]{0.65\textwidth}
		\centering
		\includegraphics[
		width=\textwidth,
		keepaspectratio
		]{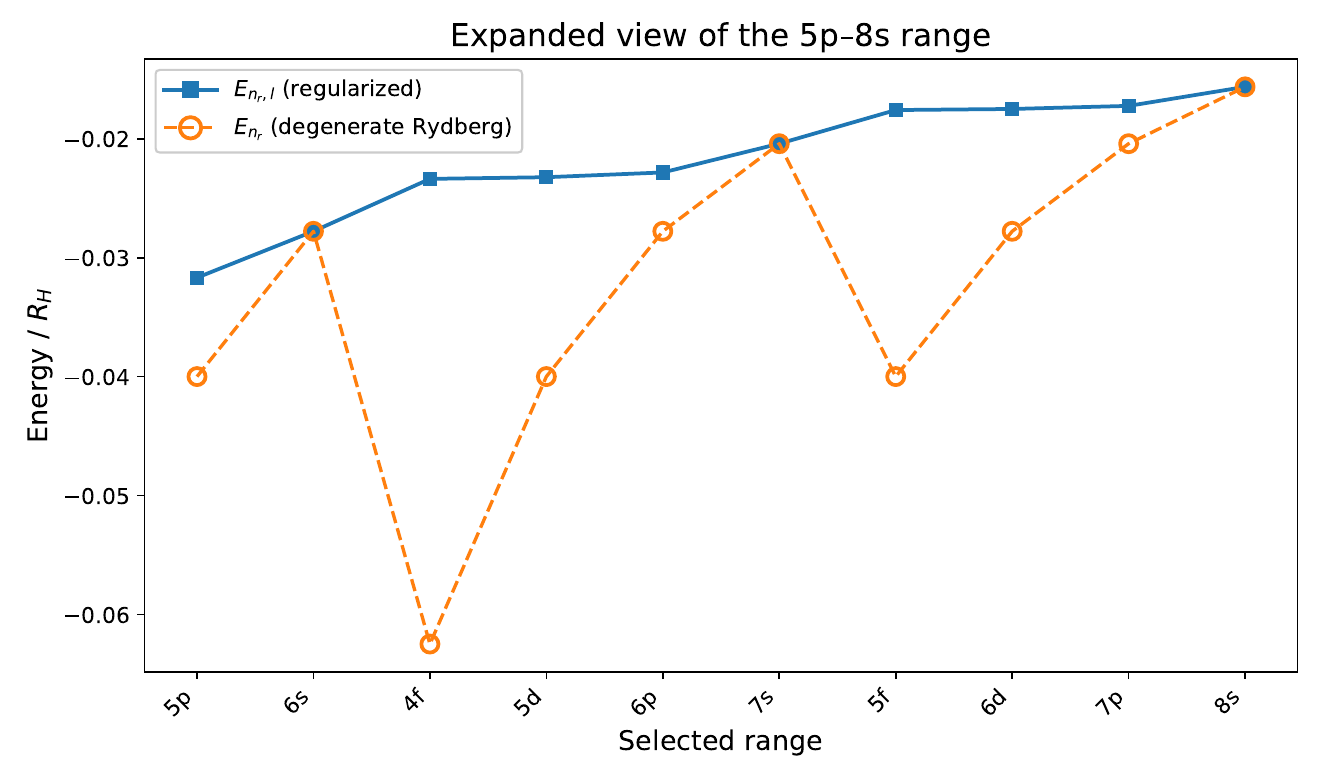}
		\caption{Expanded view of the \(5p\) -- \(8s\) range.}
		\label{fig:Aufbau_zoom_overlap}
	\end{subfigure}
	
	\caption{
		Comparison of the degenerate Rydberg energies \(E_{n_r}\) with the shifted
		regularized spectrum \(E_{n_r,l}\) along the Aufbau-ordered sequence.
		The upper panel shows the full displayed range, while the lower panel
		expands the higher-state region where the energy separation is smaller.
	}
	\label{fig:AufbauEnergy}
\end{figure}

\noindent
In the previous section, the minimum branch of $E_{n_r,\lambda}$ was identified and the separation label $\lambda$ was assigned to the orbital label $l$ through $\lambda=l-\tfrac12$. This allows the non--degenerate regularized spectrum to be expressed in the conventional ($s,p,d,f,\ldots$) notation. From Eq.~\eqref{eq:enrl}, the $s$-sector maps exactly to the hydrogenic Rydberg sequence.
\[
E_{n_r,s}=-\frac{R_H}{(n_r+1)^2},
\]
whereas higher orbital sectors retain regularization--induced fractional shifts. Thus, the $n_r,l$ branches form a non--degenerate spectrum, which, under the minimum labeling scheme, can be compared directly with the Aufbau/Madelung ordering. The corresponding regularized energies and degenerate Rydberg values are shown in Fig.~\ref{fig:AufbauEnergy}.

\vspace{0.5pc}

\noindent
Over the displayed range, the regularized bound--state energies follow the sequence

\[
\begin{array}{l}
	\boxed{
		1s < 2s < 2p < 3s < 3p < 4s < 3d < 4p < 5s < 4d < 5p < 6s < 4f < 5d < 6p < 7s < 5f < 6d < 7p < 8s < 5g.
	}
\end{array}
\]
This reproduces the Aufbau ordering over the displayed branch sequence.

\vspace{0.5pc}

\noindent 
It is useful to compare the present result with the WKB form of the Langer correction. In the notation used here, with $\lambda$ denoting the separation label, the WKB-equivalent radial expression may be written as
\[
E^{\mathrm{WKB}}_{n_r,\lambda}
=
-\frac{R_H}{
	\left(
	n_r+\frac12+\sqrt{\lambda(\lambda+1)}
	\right)^2
}.
\]
By contrast, the Langer-like correction obtained from the Sturm--Liouville reduction and the terminating confluent-hypergeometric, or associated Laguerre, solution gives
\[
E^{\mathrm{SL}}_{n_r,\lambda}
=
-\frac{R_H}{
	\left(
	n_r+\frac12+\sqrt{\lambda(\lambda+1)+\frac12}
	\right)^2
}.
\]
The distinction between these two expressions is important. In the usual Coulomb problem, the Langer modification is primarily interpreted as a WKB or uniform-asymptotic correction that restores the correct radial quantization near the second-order pole at the origin. In that setting the correction is usually applied after the orbital label has already been fixed. In the present formulation, however, the shifted inverse-square structure is visible directly at the level of the radial Sturm--Liouville equation and its associated Laguerre solution. The separation parameter \(\lambda\) is first treated as a regularized spectral label, and the terminating condition exposes the shifted radial index explicitly.
\vspace{0.5pc}

\noindent
The additional \(+\frac12\) inside the square root in \(E^{\mathrm{SL}}_{n_r,\lambda}\) is therefore not a cosmetic change. It permits a real minimum branch at \(\lambda=-1/2\), after which the comparison with hydrogenic orbital labels is made through \(\lambda=l-\frac12\). This step converts the Langer-like radial shift from a semiclassical quantization repair into a mechanism for organizing the Coulomb branches into an Aufbau-like sequence. Thus, the Aufbau-like ordering follows from the Sturm--Liouville/Laguerre realization of the Langer-like correction, rather than from the WKB expression alone.

\section{Summary and conclusions}
\label{sec:Sec5}

\noindent
We examined the stationary Coulomb problem in a regularized de Broglie--Bohm representation. Amplitude--phase decomposition, together with stationary current constraints, leads to component-wise Sturm--Liouville equations in spherical coordinates. Imposing the canonical regularization condition introduced Langer--like inverse--square contributions to the radial, orbital, and axial sectors. As a result, the separated amplitudes are no longer represented by the standard hydrogenic spherical--harmonic structure, but by the generalized radial Laguerre, associated Legendre, and Bessel branches.
\vspace{0.5pc}

\noindent
The principal result is the shifted radial spectrum
\[
E_{n_r,l}
=
-\frac{R_H}{
	\left(
	n_r+\frac12+\sqrt{l^2+\frac14}
	\right)^2},
\]
obtained after assigning the minimum spectral branch
\[
\lambda=l-\frac12,\qquad l=0,1,2,\ldots .
\]
For $l=0$, this expression reduces exactly to the degenerate Rydberg sequence,
\[
E_{n_r,s}=-\frac{R_H}{(n_r+1)^2}.
\]

\noindent
For $l>0$, the regularized spectrum lifts Coulomb degeneracy and produces a shifted orbital hierarchy. When the minimum labeling scheme is used to compare the regularized branches with the conventional ($s,p,d,f,\ldots$) notation, the resulting sequence follows the Aufbau/Madelung ordering over the displayed range.

\vspace{0.5pc}

\noindent
This construction should not be interpreted as a replacement for many--electron-atom theories. Screening, penetration, exchange, correlation, and relativistic effects remain essential for quantitative atomic structure and for known exceptions to empirical filling rules. The present result instead isolates a simpler mathematical question: can a one--center stationary Coulomb problem, when regularized through de Broglie--Bohm amplitude--current constraints, already contain  intrinsic orbital splitting? Within this restricted sense, the regularized spectrum provides an analytical mechanism by which an Aufbau--like ordering emerges from shifted Coulomb quantization.

\vspace{0.5pc}

\noindent
The angular sector also clarifies the mathematical status of the regularized amplitudes. The associated Legendre and Bessel factors are generalized non--integral special function branches, and the typical spherical--harmonic closure is no longer imposed on the separated angular representation. Therefore, the selected labels serve two roles: they define admissible regularized branches of the separated equations, and they provide a minimum correspondence with the conventional hydrogenic orbital notation. This establishes the regularized Coulomb problem as a useful analytical model for studying how orbital ordering can arise from stationary amplitude regularization.
\vspace{0.5pc}

\noindent This result is closely related to the Langer--like regularizations of radial quantum problems, where inverse--square corrections repair the WKB quantization of the Coulomb problem near the second--order pole at its origin. The distinction here is that the shift is derived from the canonical regularization of the stationary de Broglie--Bohm amplitude--phase equations and is then used to organize the Coulomb branches into an Aufbau--like spectral ordering.
\vspace{1pc}

\noindent \textbf{Acknowledgements.} The author thanks Rajesh Tengli and S.~K.~Srivatsa for their useful discussions, careful reading, and helpful comments during the preparation of this manuscript.
\vspace{0.5pc}

\noindent\textbf{Funding Statement.}
The author received no external funding for this study.

\vspace{0.5pc}

\noindent\textbf{Data Availability Statement.}
No datasets were generated or analyzed in this study.

\vspace{0.5pc}

\noindent\textbf{Ethics Statement.}
This study did not involve human participants, animals, or other identifiable personal data.

\vspace{0.5pc}

\noindent\textbf{Declaration of Interest:}
This research is an independent work of the author and declares no conflicts of interest with the employer, IBM Research, Albany, NY, USA.

\vspace{0.5pc}

\noindent\textbf{Use of Generative AI.}
Generative AI tools were used to prepare this manuscript for language editing, grammar refinement, and equation cross-verification. These tools were not used to generate the scientific results, derivations, or core interpretations presented herein. All mathematical checks, technical conclusions, and final editorial decisions are the sole responsibilities of the author.

\end{document}